\documentclass[
]{ceurart}

\usepackage{graphicx} 
\usepackage{caption} 

\usepackage{listings}
\usepackage{threeparttable}
\usepackage{subcaption}
\begin{document}

\copyrightyear{2025}
\copyrightclause{Copyright for this paper by its authors.
  Use permitted under Creative Commons License Attribution 4.0
  International (CC BY 4.0).}

\conference{Transition Network Analysis Workshop (TNA 2026), 
co-located with LAK 2026,
April 27--28, 2026, Bergen, Norway}

\title{Towards Modeling Situational Awareness Through Visual Attention in Clinical Simulations}

\author[1]{Haoting Gao}[
orcid=0009-0002-8159-1975,
email=haotin@umich.edu
]\cormark[1]

\author[1,2]{Kapotaksha Das}[
orcid=0000-0001-9920-4668,
email=takposha@umich.edu,
]\cormark[1]

\author[2]{Mohamed Abouelenien}[
orcid= 0000-0001-5351-5778,
email= zmohamed@umich.edu,
]

\author[1]{Michael Cole}[
orcid= 0000-0000-0000-0000,
email= mcolemd@umich.edu,
]

\author[1]{James Cooke}[
orcid= 0000-0000-0000-0000,
email= cookej@umich.edu,
]

\author[1]{Vitaliy Popov}[
orcid= 0000-0003-2348-5285,
email= vipopov@umich.edu,
]

\address[1]{Department of Learning Health Sciences, 
  University of Michigan Medical School, 
  Ann Arbor, MI, United States}

\address[2]{Computer and Information Sciences, 
  University of Michigan-Dearborn, 
  Dearborn, MI, United States}

\cortext[1]{Corresponding author.}

\begin{abstract}
Situational awareness (SA) is essential for effective team performance in time-critical clinical environments, yet its dynamic and distributed nature remains difficult to characterize. In this preliminary study, we apply Transition Network Analysis (TNA) to model visual attention in multiperson VR-based cardiac arrest simulations. Using eye-tracking data from 40 clinicians assigned to four standardized roles (Airway, CPR, Defib, TeamLead), we construct gaze transition networks between clinically meaningful areas of interest (AOIs) and extract metrics such as entropy and self-loop rate to quantify attentional structure and flow. Our findings reveal that individual and team's visual attention is dynamically and adaptively redistributed across roles and scenario phases, with those in CPR roles narrowing their focus to execution-critical tasks and those in the TeamLead role concentrating on global monitoring as clinical demands evolve. TNA thus provides a powerful lens for mapping functional differentiation of team cognition and may support the development of phase-sensitive analytics and targeted instructional interventions in acute care training.
\end{abstract}


\maketitle

\footnote{This paper has been accepted to the Workshop on Transition Network Analysis (TNA), 
co-located with the Learning Analytics and Knowledge Conference (LAK 2026). 
This arXiv version is a preprint.}

\section{Introduction}
Understanding how teams coordinate and maintain situational awareness in time-critical, high-stakes environments remains a challenge in behavioral and learning analytics \cite{popov2025communication, sebok2025we,keller2024behavioral}. Multi-disciplinary research on acute care teams has long sought to characterize team performance through observational approaches, primarily behavioral marker systems and coding schemes \cite{kolbe2019laborious}. However, these methods are labor-intensive, obtrusive, prone to subjective judgment, and fundamentally unable to capture the dynamic, interdependent, and multimodal nature of team processes as they unfold in real time \cite{kolbe2019laborious, lavoie2023comparison}. There is a need for scalable, objective, and temporally sensitive analytical approaches that can model collaborative problem-solving as it naturally occurs.

Recent advances in sensor technology and computational modeling offer a promising alternative to traditional approaches (e.g., self-report, observation, etc.) \cite{rosen2015integrative}. High-fidelity data streams, including visual attention (gaze), speech patterns, physiological arousal, and behavioral sequences, can now be captured unobtrusively during simulated training or real-world clinical performance \cite{rosen2015integrative}. Among these data modalities, gaze-based fixation data can serve as a particularly valuable proxy for situational awareness (SA) \cite{caloca2024exploring,lavoie2023comparison}. Visual attention reflects how individuals perceive environmental cues, monitor task progress, and coordinate with team members \cite{weinberg2020visual, li2023using}. Yet traditional gaze metrics, such as fixation counts, dwell times, or pairwise area-of-interest (AOI) transitions, fail to capture the temporal structure and sequential reorganization of attention across evolving task demands and across multiple team members simultaneously. 

To address this gap, we leverage Transition Network Analysis (TNA) \cite{saqr2025transition}, a graph-based approach that models visual attention as a temporal network of gaze transitions between clinically meaningful AOIs. TNA moves beyond where clinicians look to examine how attention flows and adapts across roles and scenario phases. By integrating network-level metrics such as entropy (attentional dispersion) and self-loop rate (sustained focus), TNA offers a clear vantage on latent structures in collaborative visual attention. 

We situate our analysis within multiperson cardiac arrest virtual reality (VR) simulations, where improved SA and teamwork can significantly affect patient outcomes \cite{kentros2025non, nallamothu2018resuscitation, holmberg2019trends}. While VR environments may differ from real-world settings in how visual attention is allocated (e.g., altered depth cues, peripheral awareness, and field of view), VR uniquely affords an immersive, repeatable (controlled for stimulus presentation) platform for capturing rich, multimodal interactions. Traditional training and assessment methods provide limited systematic insight into individual and team SA during acute events \cite{popov2025elucidating, trevi2024virtual}. However, a deeper understanding of how attentional structures differ by clinical role and evolve across scenario stages remains lacking.

In this study, we present a preliminary application of TNA to multi-role, real-time visual attention in VR-based cardiac arrest simulations. We ask: \textit{How can situational awareness be dynamically modeled using visual attention patterns across different clinical roles and scenario stages in multiperson cardiac arrest simulations?} 

Using eye-tracking fixation data, we construct role-specific gaze transition networks, quantify their structure via entropy and self-loop rate, comparing patterns across roles and scenario stages.

\section{Methodology}
\subsection{Study Context and Data Collection}
This study utilized an interactive, real-time VR platform designed for medical emergency training. The simulation provides a flexible, nonlinear environment in which teams of four medical trainees collaborate to manage rapid-response hospital scenarios, with a focus on cardiac arrest and acute dysrhythmias. Each 10-15 minute session assigned trainees to standardized roles (TeamLead, Airway, CPR, Defib), following American Heart Association guidelines to collaboratively assess and respond to evolving clinical situations \cite{perman20242023}. 
The simulation tracks a STEMI cardiac arrest scenario \cite{thygesen2018fourth}, where a blocked coronary artery triggers life-threatening arrhythmias. The patient presented with ventricular tachycardia and a weak pulse (Stage 1), progressed through pulseless VT, asystole, and ventricular fibrillation (Stage 5), before return of circulation (ROSC). Stages 1 and 5 were chosen for analysis (see Sec. \ref{sec:stageRoleAOI}) as they represent initial assessment and peak intervention phases, capturing major shifts in team attention and coordination.

For this exploratory research, data were collected from 40 participants across 10 sessions. The teams had varying skill levels, ranging from experienced clinicians such as paramedics and physicians to junior emergency and family medicine residents with Advanced Cardiovascular Life Support (ACLS) training but limited real-world leadership experience in cardiac arrest scenarios. Using a VR platform with HP Reverb G2 Omnicept headsets, in conjunction with the Cognitive3D \cite{Cognitive3D} software, enabled the collection and processing of rich movement, action, and eye-tracking data, yielding fixation data for each participant in the simulation. All data collection associated with these sessions was approved by the Institutional Review Board (IRB).

Simulations had an average duration of 11.66 minutes (SD = 3.65), yielding a dense set of eye-tracking data, including 55{,}670 raw fixations and 22{,}660 saccades. After merging adjacent fixations on the same object with inter-fixation gaps $\leq 300$~ms, the resulting dataset comprised 20{,}628 merged fixations and 20{,}526 Areas of Interest (AOI) transitions, which served as the basis for all transition matrices and TNA metrics.

\begin{table}[b]
\centering
\caption{Areas of Interest (AOIs) Used for Gaze–Transition Analysis}
\label{tab:AOIs}
    \begin{tabular}{cl}
    \hline
    \textbf{AOI}           & \textbf{Description}                             \\ \hline
    Equipment - Airway     & Airway management equipment, such as bag-valve masks.         \\
    Equipment - CPR        & Hands and UI elements directly involved in CPR.  \\
    Equipment - Defib      & Main Defibrillator unit and associated controls. \\
    Equipment - Meds \& IV & Medication- and IV-related objects.              \\
    Patient Vitals Monitor & Displays patient vitals such as ECG, SpO2, Respiration, etc. \\
    Other Team Members     & 3D avatars of other people in session.           \\
    Patient                & Virtual patient body.                            \\ \hline
    \end{tabular}
\end{table}

\begin{figure}[t]    
\centering
    \includegraphics[width=0.43\textwidth]{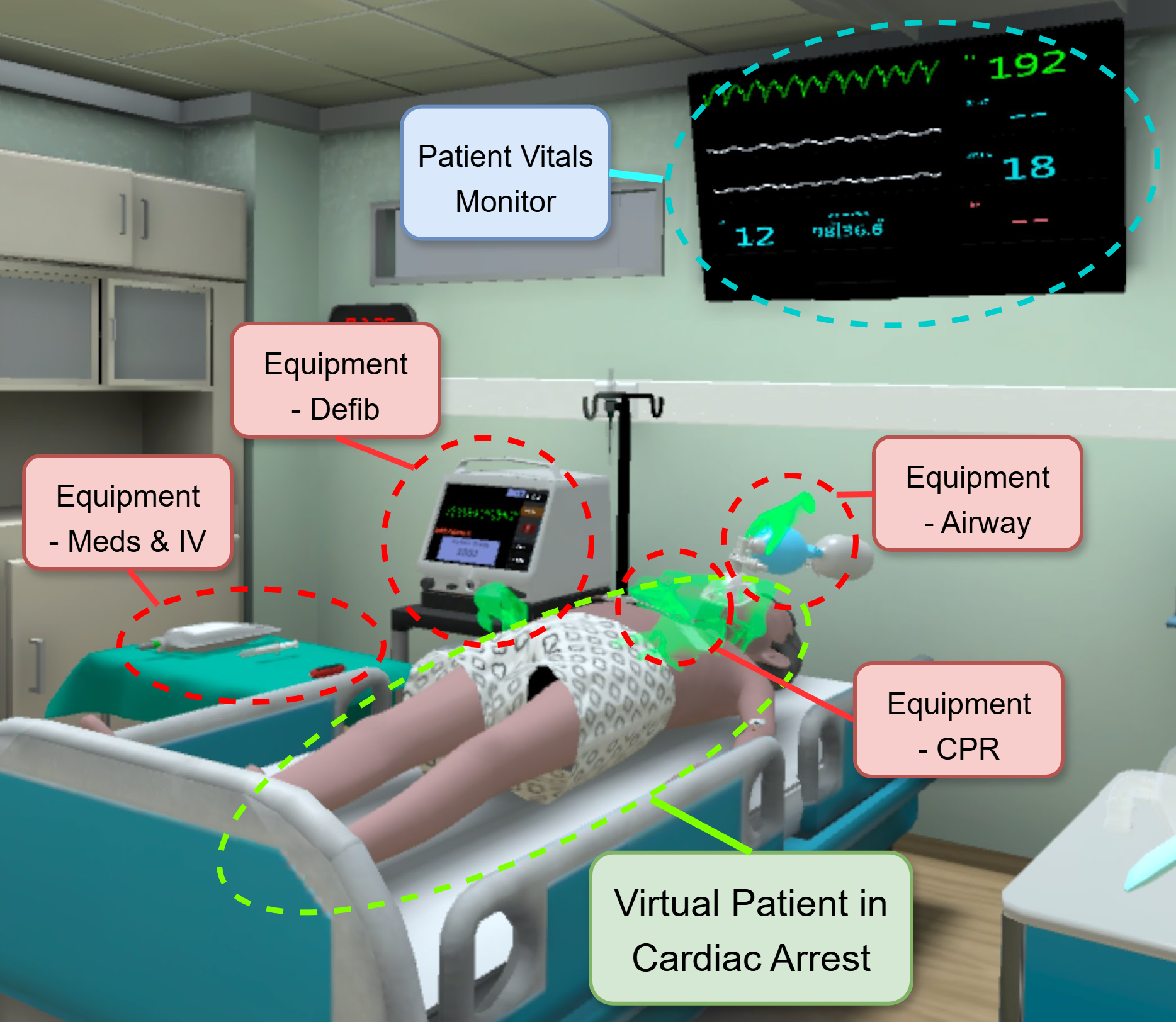}
    \captionof{figure}{Visual for AOIs as seen in VR Environment ("Other Team Members" AOI not present in this Figure).}
    \label{fig:AOIs}
    \vspace{-8pt}
\end{figure}

\subsection{Defining Areas of Interest (AOIs)}
\label{sec:AOIS}
To understand participants' gaze patterns during the VR cardiac scenario, we defined seven Areas of Interest (AOIs) that served as perceptual anchors in the simulation. These AOIs group objects by key roles or use cases, such as airway management, CPR, defibrillation, and medication, as well as by patient, vital sign monitors, and avatars representing other team members. Table \ref{tab:AOIs} lists the seven AOIs used, along with a description of the objects included in each category, with Fig. \ref{fig:AOIs} showing a visualization of what the objects look like in the VR environment.

\subsection{Gaze Metrics and Transition Construction}
To characterize the dynamic structure of clinicians' visual attention during cardiac arrest, we developed an analytical framework that models gaze-transition networks and their progression across clinically defined scenario stages. Visual attention forms the foundation of Level 1 SA in Endsley's model~\cite{ensley1995toward} and, under stress, often narrows toward dominant or familiar information sources while peripheral cues are neglected~\cite{hockey1970effect}. This cognitive tunnel vision motivates our focus on both exploratory transitions and sustained self-loops in the subsequent TNA metrics. To capture these attentional dynamics, we transformed clinicians’ eye fixations into transitions between AOIs in the simulation. We constructed role-specific AOI fixation sequences, estimated transition probability matrices, and extracted TNA metrics (entropy, self-loop rate).

\subsubsection{AOI Fixation Sequence Construction}
Eye-tracking data were grouped by each clinician's assigned role (Airway, CPR, Defib, Teamlead). Using the AOIs defined in Sec. \ref{sec:AOIS}, every fixation instance was mapped and ordered by timestamp to form a role-specific AOI fixation sequence. Fixations without valid AOI labels and all saccades were removed. Adjacent AOI pairs (AOI$_i \rightarrow$ AOI$_j$) were extracted to represent moment-to-moment attentional transitions. These directed transitions serve as the basis for constructing the transition networks.

\subsubsection{Transition Matrix Estimation}
For each participant-role, we computed a $7 \times 7$ AOI transition count matrix $C$, where $C_{ij}$ denotes the number of transitions from AOI $i$ to AOI $j$. To stabilize transition probability estimates for sparsely populated rows, we applied mild Laplace smoothing ($\alpha = 0.5$) to all non-empty rows. The smoothed transition probability matrix $P$ was obtained by row normalization:
\vspace{-4pt}
\[
P_{ij} = \frac{C_{ij} + \alpha}{\sum_{k}(C_{ik} + \alpha)},
\]
producing a row-stochastic matrix that models each clinician's attentional routing structure.

\subsubsection{Transition-Network Metrics}
Treating $P$ as a weighted directed graph, we computed two complementary TNA-based metrics corresponding to key components of SA.
\paragraph{Entropy}
Entropy quantifies the unpredictability or dispersion of outgoing transitions from each AOI. We computed row-wise entropy for each AOI $i$, then averaged across all $N = 7$ AOIs:
\vspace{-4pt}
\[
H_i = -\sum_{j=1}^{N} P_{ij}\log_2 P_{ij}, \qquad 
H = \frac{1}{N}\sum_{i=1}^{N} H_i.
\]
Prior transition-based gaze studies have typically included self-transitions when computing transition entropy  \cite{krejtz2015gaze}. However, because self-loops in our context reflect attentional tunneling (i.e., sustained fixation on a single AOI) rather than exploratory visual sampling, we computed entropy over non-self transitions only ($i \neq j$). This separation allows entropy to reflect cross-AOI unpredictability, while the self-loop metric captures attentional tunneling. 
Higher entropy indicates more diverse and exploratory visual scanning, and lower entropy reflects more structured and predictable attentional routing \cite{cui2024gaze, krejtz2015gaze}. 

\paragraph{Self-loop Rate}
Self-loop rate quantifies repeated fixations on the same AOI and reflects the extent to which attention remains anchored to specific perceptual cues, rather than shifting across different areas. It is calculated by weighting self-loops by the proportion of total fixations on each AOI ($w_i$):
\vspace{-4pt}
\[\text{Self-loop} = \sum_{i=1}^{N} w_i \cdot P_{ii}.\]

A high self-loop rate indicates strong attentional anchoring, or tunneling, in which individuals repeatedly focus on familiar or dominant cues, such as a particular monitor or piece of equipment, while neglecting surrounding information. This pattern is consistent with classical descriptions of ``attention narrowness" and ``cognitive tunnel vision," which are often observed under pressure or high task load \cite{duncan1996information, rasmussen2013human}.  Conversely, a low self-loop rate suggests a broader, more distributed pattern of visual attention, with gaze shifting more frequently between AOIs. In this case, the \emph{cross-scan rate}, which represents transitions across AOIs, is higher, indicating more active information gathering and monitoring across patient, equipment, monitors, and team members \cite{weinberg2020visual,hunziker2011teamwork}. 

Together, these TNA-derived gaze metrics provide a quantitative characterization of how clinical roles dynamically allocate visual attention during high-fidelity resuscitation simulations.

\subsection{Transition Network Analysis}

In addition to the quantitative metrics described above, we performed a qualitative inspection of the role- and stage-specific transition networks to identify salient triadic and dyadic patterns. 
Following prior TNA work \cite{saqr2025transition}, we use the term \emph{triadic motif} descriptively to refer to visually evident three-node closed loops in the AOI graphs (e.g., \textit{Equipment - CPR | Patient | Other Team Members}) and use \emph{dyadic motif} to refer to strongly recurring bidirectional AOI pairs that form stable two-node coupling structures (e.g., \textit{Equipment - Defib | Patient Vitals Monitor}). 
These motifs are treated as interpretive units that summarize how two and three AOIs tend to be co-visited as part of recurring attentional routines, rather than as formally tested structures that occur more frequently than chance. We did not apply algorithmic motif enumeration or statistical over-representation tests; instead, motif-based descriptions in Section \ref{sec:results} are grounded in repeated visual patterns observed across the aggregated role- and stage-specific networks.

\section{Results \& Discussion}
\label{sec:results}

\subsection{Role-Based Patterns in Visual Attention}

\begin{table}[t]
\centering
\caption{Comparison of Transition Network Analysis (TNA) Metrics Across Roles (Values reported as median (Q1–Q3)}
\label{tab:tableTNAmetrics}

\begin{threeparttable}

\begin{tabular}{lcccc c}
\toprule
\textbf{TNA Metric} & \textbf{Airway} & \textbf{CPR} & \textbf{Defib} & \textbf{TeamLead} & \textbf{KW $p$} \\
\midrule

Entropy &
2.06 (1.792–2.126) &
2.21 (2.182–2.241) &
2.02 (1.929–2.073) &
1.95 (1.746–2.098) &
\textbf{0.0015**}\\

Self-loop&
0.38 (0.362–0.451) &
0.39 (0.377–0.409) &
0.49 (0.479–0.525) &
0.45 (0.394–0.460) &
\textbf{0.0016**}\\

\bottomrule
\end{tabular}


\end{threeparttable}
\end{table}

\begin{figure}[t]
    \centering

    \begin{subfigure}[b]{0.38\textwidth}
        \centering
        \includegraphics[width=\linewidth]{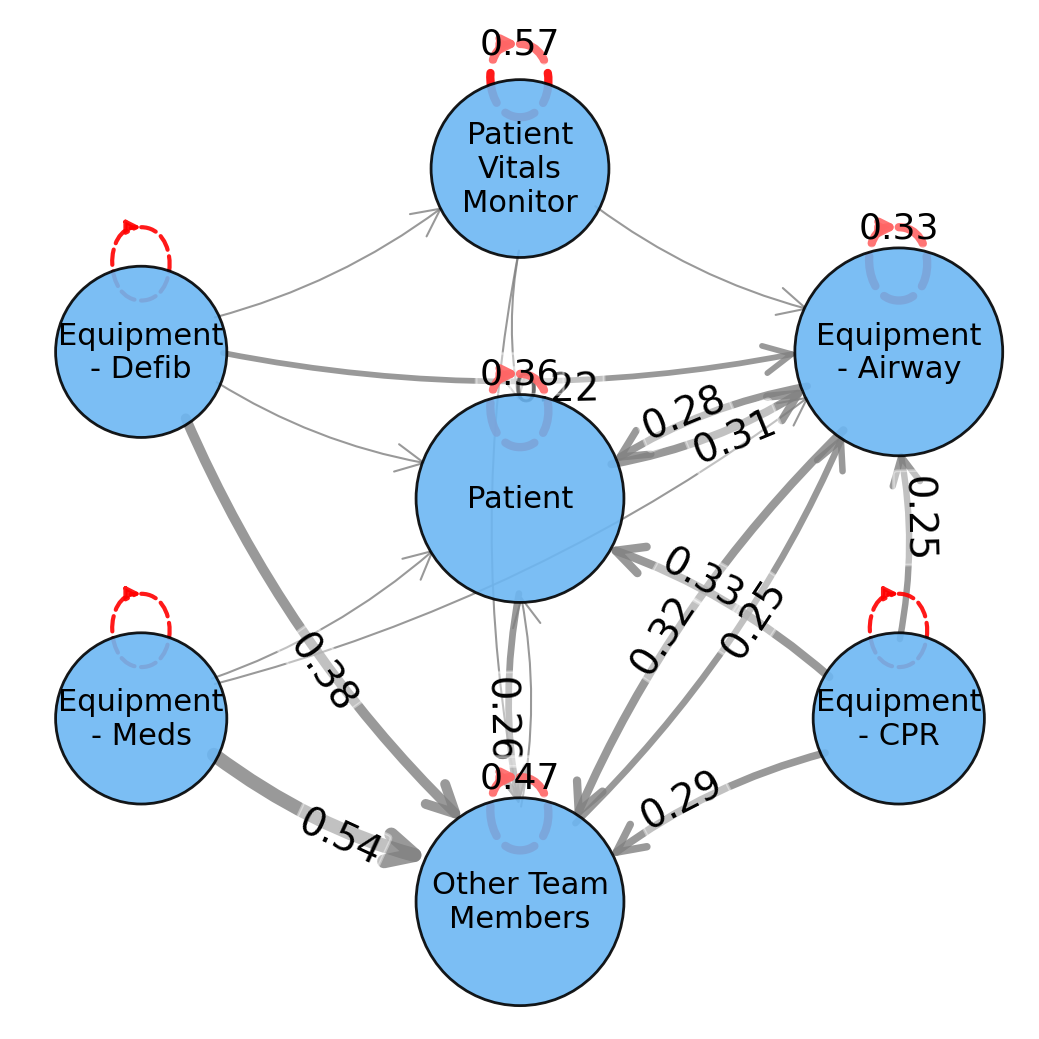}
        \caption{Airway Role}
    \end{subfigure}
    \begin{subfigure}[b]{0.38\textwidth}
        \centering
        \includegraphics[width=\linewidth]{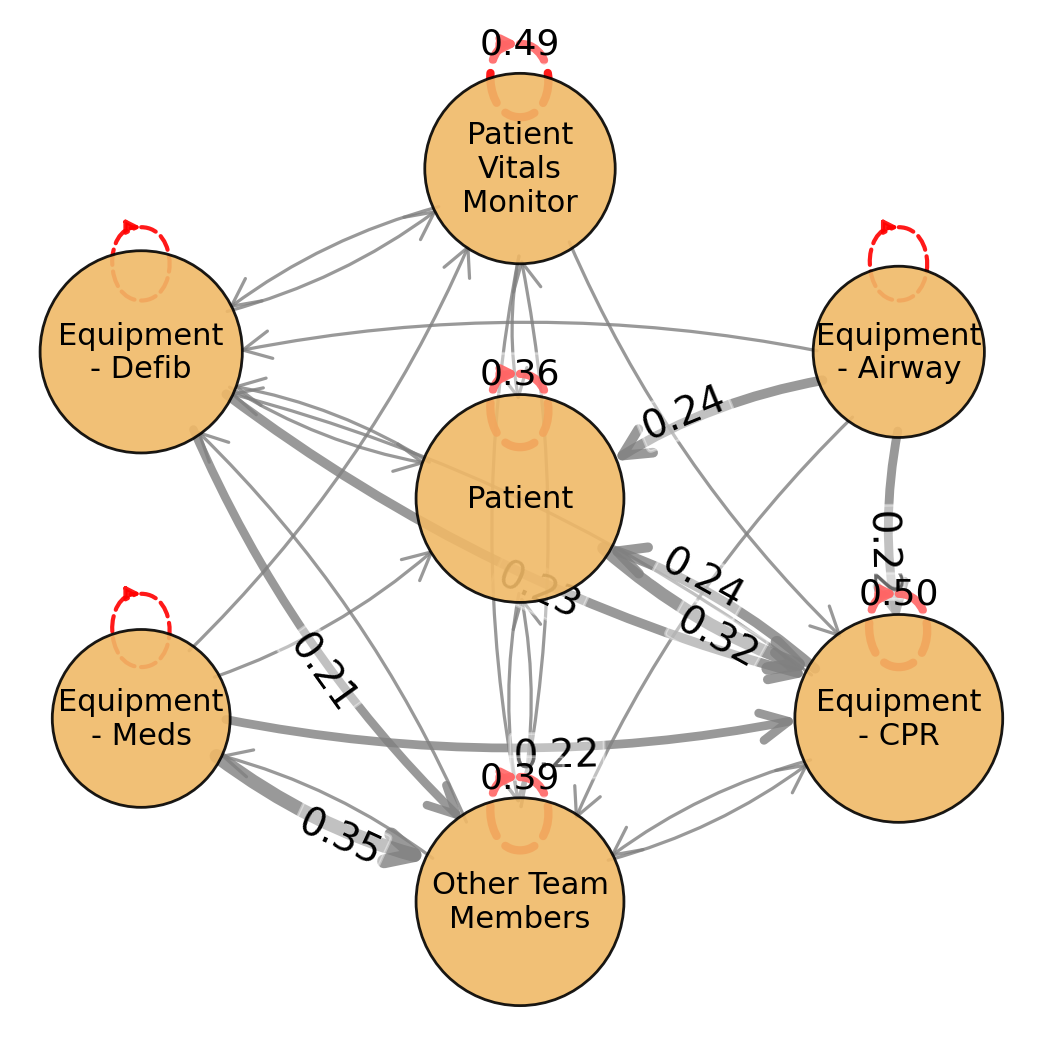}
        \caption{CPR Role}
    \end{subfigure}

    \vspace{0pt}

    \begin{subfigure}[b]{0.38\textwidth}
        \centering
        \includegraphics[width=\linewidth]{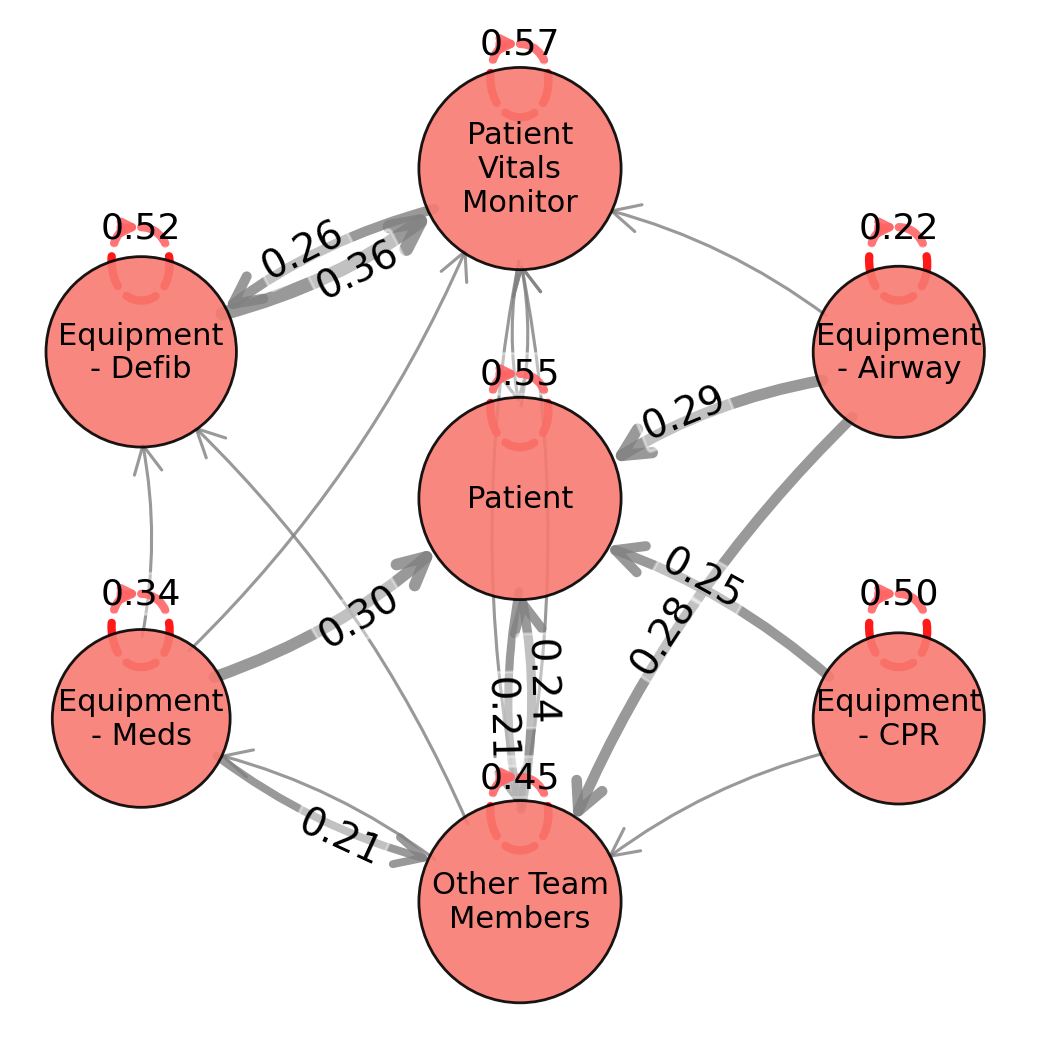}
        \caption{Defib Role}
    \end{subfigure}
    \begin{subfigure}[b]{0.38\textwidth}
        \centering
        \includegraphics[width=\linewidth]{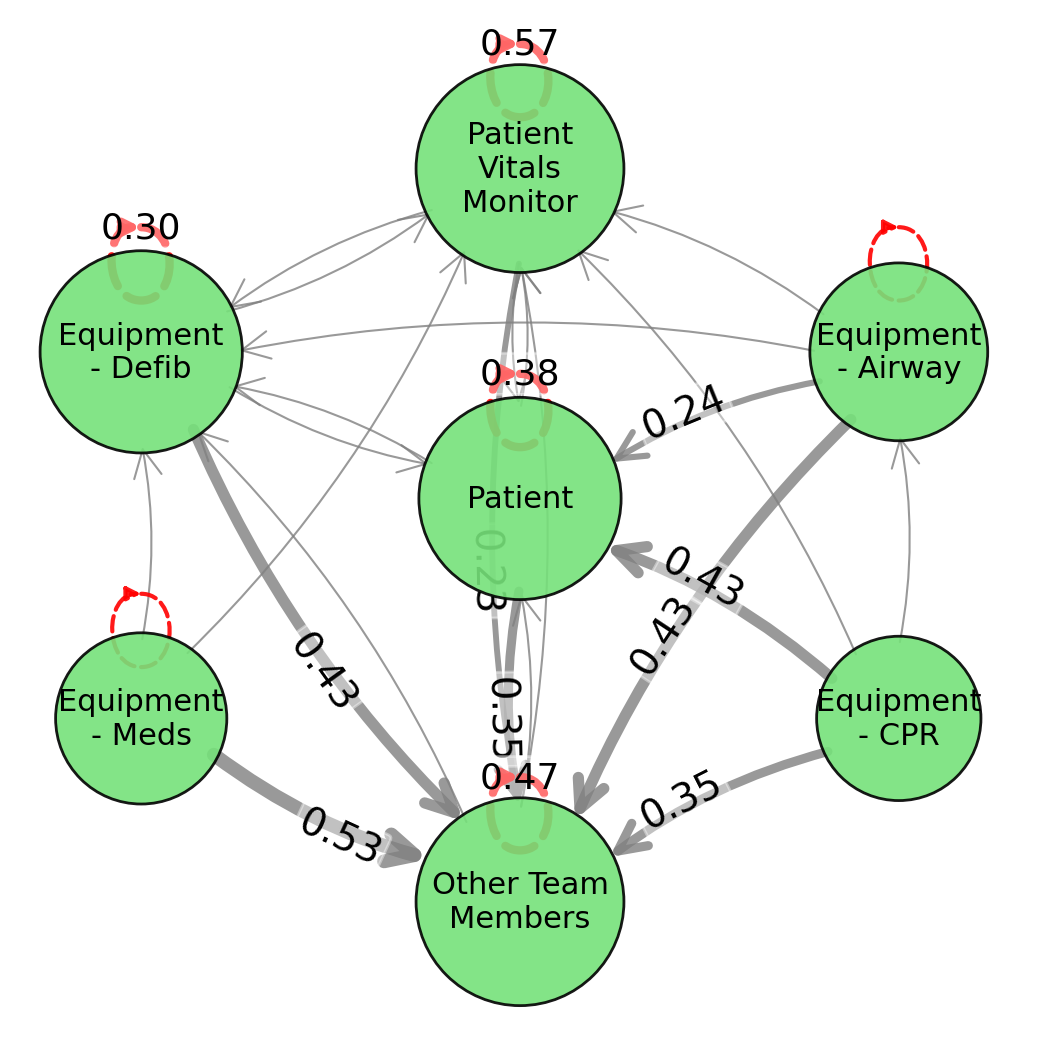}
        \caption{TeamLead Role}
    \end{subfigure}

    \caption{Role-specific AOI transition networks for four resuscitation roles across seven AOIs, aggregated over all sessions. Node size reflects total fixations; edge thickness indicates transition probability; red edges denote self-loops.}
    \label{fig:role_tna_2x2}
    \vspace{-8pt}
\end{figure}

Although gaze activity volumes were comparable across four roles (all Kruskal–Wallis $p > .47$), AOI transition network structures diverged substantially, reflecting distinct demands and information needs. This reinforces prior work indicating that members of acute care teams allocate visual attention in role-specific ways, shaped by their distinct situational demands, coordination, and information needs \cite{kolbe2013co, weinberg2020visual, white2018getting}.

Quantitative analysis of TNA metrics revealed significant role-based differences in transition entropy and self-loop rate ($p \approx .0015$). Those in CPR roles exhibited the highest transition entropy and relatively lower self-loop rates, whereas those in Defib roles showed the highest self-loop rates. Airway and TeamLead roles occupied intermediate positions on these metrics as seen in Table \ref{tab:tableTNAmetrics}.  Notably, while both CPR and Defib clinicians operated within highly centralized networks anchored by loops between their task-critical AOIs, such as \textit{Equipment - CPR | Patient} and \textit{Equipment - Defib | Patient Vitals Monitor}, the divergence in underlying gaze patterns indicates that CPR roles favored flexible multi-cue monitoring, while Defib roles showed more device-centric attentional tunneling. This aligns with prior research showing that technical roles in resuscitation often concentrate attention on immediate procedural and device cues as cognitive demands rise \cite{feller2023situational, rasmussen1982model, lavoie2023comparison, caloca2024exploring}.

Conversely, Airway and TeamLead clinicians exhibited more distributed networks, with gaze transitions spanning \textit{Equipment - Airway}, \textit{Patient Vitals Monitor}, \textit{Patient}, and \textit{Other Team Members}. Such patterns suggest more exploratory and integrative monitoring, characteristic of teamwork and adaptive coordination \cite{hunziker2011teamwork, keller2024behavioral}. Those in the TeamLead role, in particular, bridged patient data, monitors, and team dynamics, reflecting a broad situational overview consistent with effective leadership \cite{ rosenman2016assessing, nallamothu2018resuscitation}.

Qualitative inspection of role-specific transition networks (Fig. \ref{fig:role_tna_2x2}) revealed recurring dyadic and triadic gaze motifs, such as CPR clinicians cycling attention between \textit{Equipment - CPR}, \textit{Patient}, and \textit{Other Team Members}, and TeamLeads linking \textit{Patient Vitals Monitor}, \textit{Patient}, and \textit{Other Team Members} in closed loops. These interpretive patterns mirror the micro-coordination strategies described in gaze-based situational awareness (SA) studies, where clinicians integrate perceptual cues with team interactions to maintain continuity of action \cite{weinberg2020visual, caloca2024exploring}. Through the lens of Endsley's multi-level SA model \cite{ensley1995toward}, these motifs illustrate how roles support both Level 1 SA (cue perception) and Level 2 SA (meaning integration). In this framework, CPR clinicians couple patient status and feedback to take immediate action, and TeamLeads integrate information from multiple sources to provide a bird's-eye view and global oversight.

Altogether, these role-based analyses demonstrate that SA in acute cardiac resuscitation is encoded in the structure and dynamics of visual attention networks—distinctly shaped by role responsibilities, and emergent through both focused and distributed gaze behaviors that reflect task demands and teamwork situational imperatives.

\subsection{Stage-Based Shifts in Attentional Structure}
\label{sec:stageRoleAOI}

\begin{figure}[ht]
    \centering

    \begin{subfigure}[b]{0.38\textwidth}
        \centering
        \includegraphics[width=\linewidth]{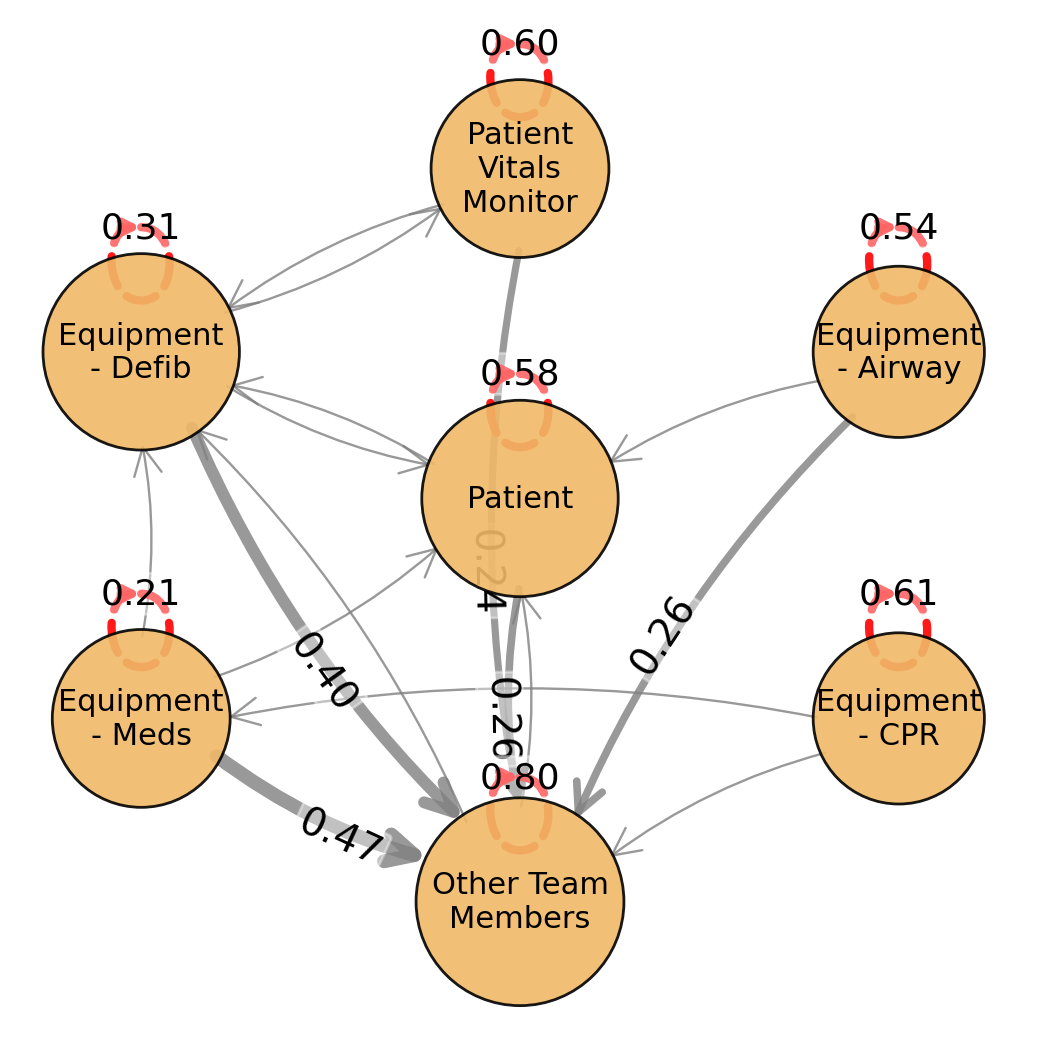}
        \caption{CPR Role - Initial Simulation Stage}
        \label{fig:cpr1}
    \end{subfigure}
    \begin{subfigure}[b]{0.38\textwidth}
        \centering
        \includegraphics[width=\linewidth]{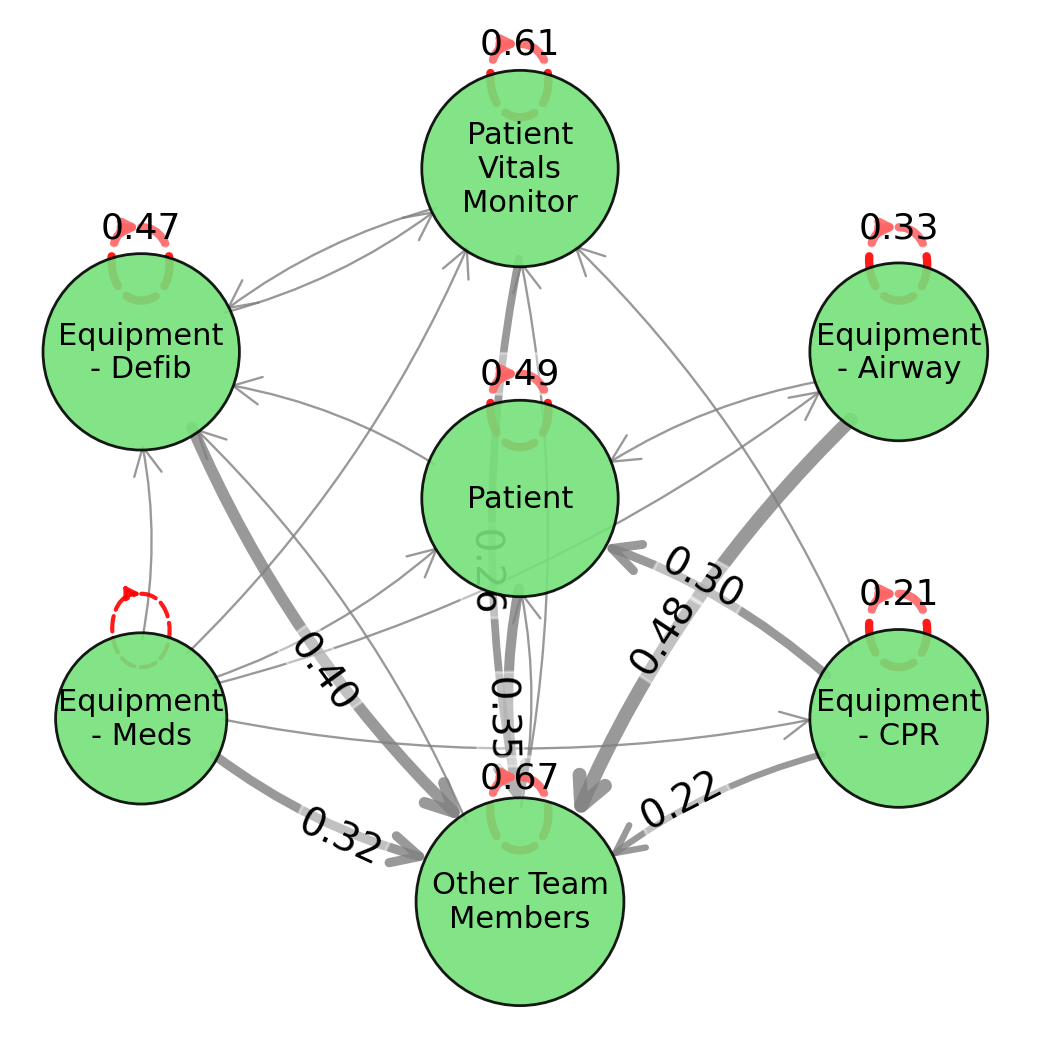}
        \caption{TeamLead Role - Initial Simulation Stage}
        \label{fig:tl1}
    \end{subfigure}

    \vspace{0pt}

    \begin{subfigure}[b]{0.4\textwidth}
        \centering
        \includegraphics[width=\linewidth]{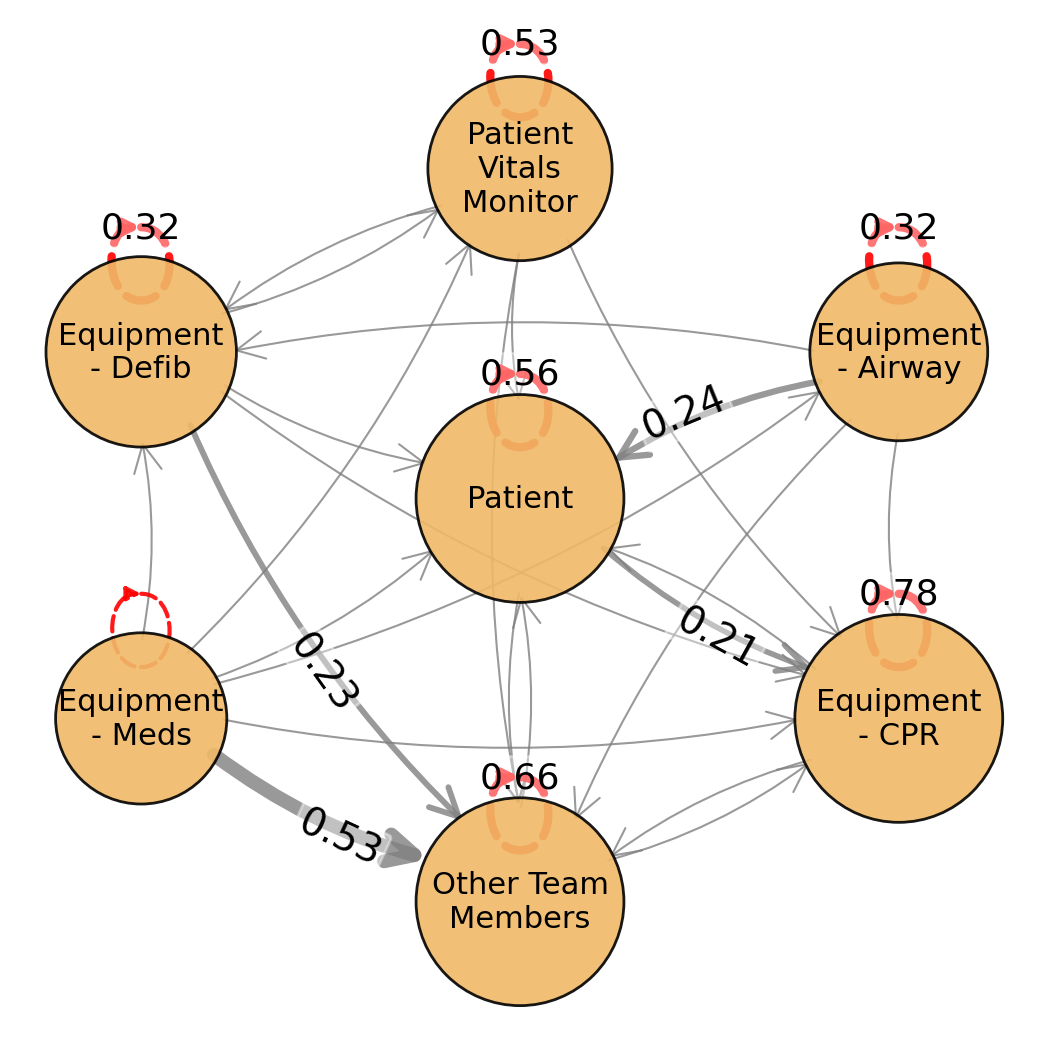}
        \caption{CPR Role - Later-Intervention Stage}
        \label{fig:cpr5}
    \end{subfigure}
    \begin{subfigure}[b]{0.4\textwidth}
        \centering
        \includegraphics[width=\linewidth]{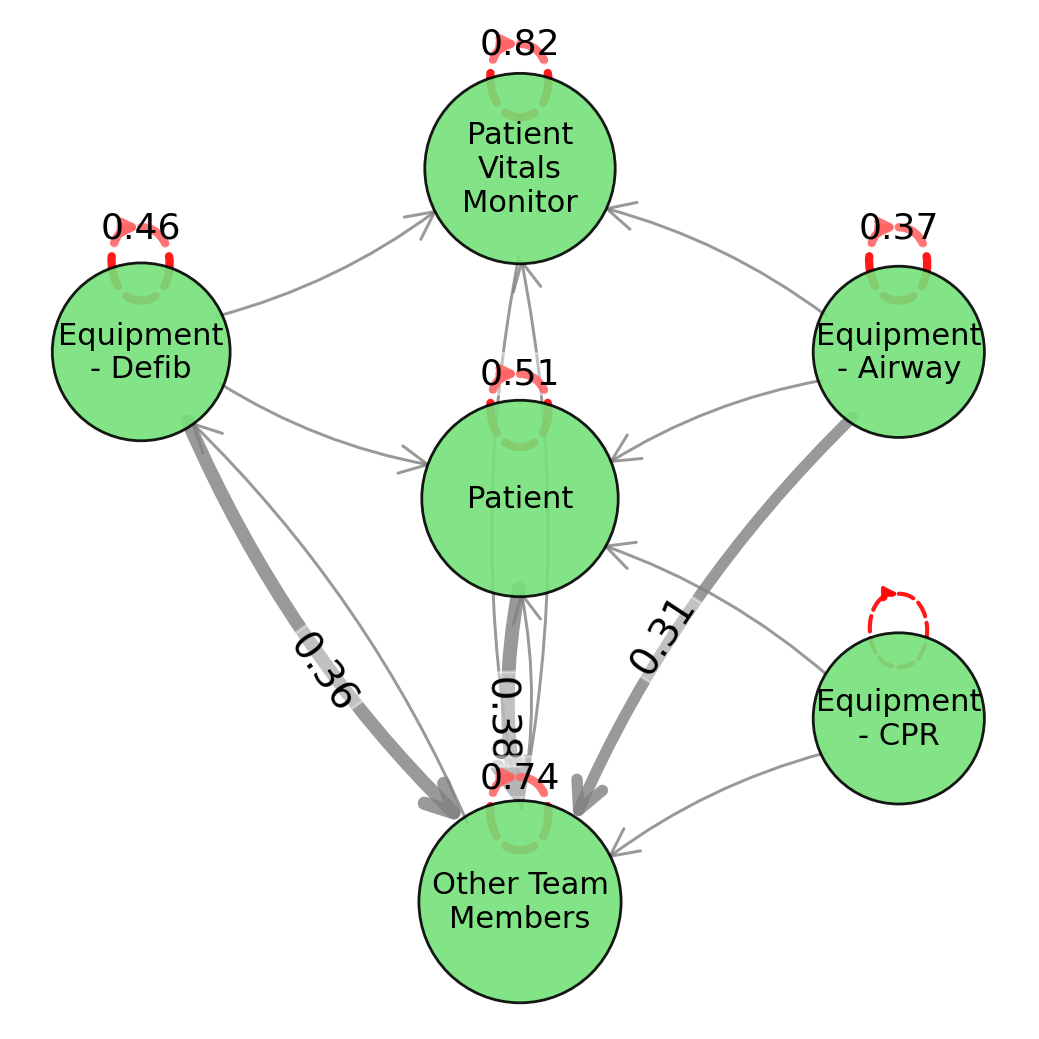}
        \caption{TeamLead Role - Later-Intervention Stage}
        \label{fig:tl5}
    \end{subfigure}

    \caption{Transition Network Diagrams for CPR and TeamLead Roles across two Stages of the Cardiac Arrest Simulation. Fig. \ref{fig:cpr1} and \ref{fig:tl1} from Stage 1, are at the start of the simulation, while Fig. \ref{fig:cpr5} and \ref{fig:tl5} are from Stage 5, later in the simulation.}
    \label{fig:two_stage_compare}
    \vspace{-8pt}
\end{figure}

We focus on the CPR and TeamLead roles as they exemplify contrasting attentional strategies, with the CPR role expected to deliver hands-on chest compressions and track immediate patient response, while the TeamLead oversees team coordination, diagnoses, and decides on intervention steps. At the start of the simulation (Stage 1), clinicians in the CPR role showed broad exploratory scanning, distributing attention across multiple AOIs, including \textit{Equipment - Defib}, \textit{Patient}, and \textit{Other Team Members}, with substantial self-loops on \textit{Other Team Members} ($.80$), \textit{Equipment - CPR} ($.61$), and \textit{Patient Vitals Monitor} ($.60$). This pattern reflects the early diagnostic demands of rhythm confirmation, coordination with teammates, and integration of diverse perceptual cues before chest compression. As the scenario progressed to later intervention phases (Stage 5), the CPR gaze network became markedly more centralized around \textit{Patient} and \textit{Equipment - CPR}; self-looping on \textit{Equipment - CPR} rose, while self-loops on \textit{Other Team Members} and \textit{Patient Vitals Monitor} decreased. This shift is consistent with increased motor workload during compressions, as attention is adaptively drawn to execution-critical AOIs and peripheral scanning diminishes—a phenomenon well documented in both perceptual psychology and acute resuscitation research \cite{hockey1970effect, duncan1996information, white2018getting}. Narrowing attentional focus is therefore adaptive, enabling CPR providers to maintain output quality while situational monitoring responsibilities shift to other team members \cite{hunziker2011teamwork, nallamothu2018resuscitation}.

In contrast, TeamLead participants followed a different trajectory. Their initial scanning focused on broad triadic motifs, integrating \textit{Patient Vitals Monitor}, \textit{Equipment - Defib}, and \textit{Other Team Members}, with moderate self-loops at key anchors. This pattern highlights the leader’s early responsibility for synthesizing information and stabilizing team workflow. By the intervention phase, TeamLead attention shifted toward monitoring-centric oversight: self-loops increased sharply on the \textit{Patient Vitals Monitor} ($.61$ → $.82$) and \textit{Other Team Members} ($.67$ → $.74$), while cross-node transitions declined. Fig. \ref{fig:two_stage_compare} illustrates these stage-wise changes in attentional network structure for both CPR and TeamLead roles. Rather than indicating reduced situational awareness, this reflects an optimized supervisory strategy—leaders prioritize temporal coordination and team collaboration under high task intensity, assuming greater responsibility for global monitoring as execution-focused roles become saturated~\cite{nallamothu2018resuscitation, rosenman2016assessing, keller2024behavioral}. These stage-dependent shifts underscore how team cognition dynamically redistributes situational awareness in response to evolving clinical demands.

Taken together, the stage-wise shifts observed in CPR and TeamLead roles reveal a coordinated, functional reorganization of situational awareness (SA) across the resuscitation team. The divergent patterns observed illustrate the functional differentiation of SA, in which cognitive responsibilities are flexibly partitioned across roles in response to changing task demands. As CPR providers' attentional bandwidth contracts around execution, TeamLeads expand and stabilize their monitoring to maintain global team awareness—a compensatory strategy consistent with distributed cognition theories in acute care \cite{keller2024behavioral}. Importantly, TNA-derived motifs provide structural evidence for this redistribution: CPR networks converge into execution dyads, while TeamLead networks organize into high-stability self-loops. Overall, these findings demonstrate that SA in cardiac arrest resuscitation is not uniformly shared, but instead is dynamically rebalanced across roles and scenario stages to support coordinated performance under stress. 

\section{Conclusion \& Future Work}

This study demonstrates how Transition Network Analysis (TNA) reveals the dynamic and compensatory nature of situational awareness in multiperson cardiac arrest VR simulations. TNA reveals coordinated, adaptive redistribution of situational awareness among team roles as scenarios unfold. CPR focus narrows during execution, while TeamLeads expand monitoring to support collective performance. Our findings highlight that situational awareness is neither static nor equally shared, but is dynamically reorganized across team roles, scenario phases, and clinical events. Recurring gaze patterns illustrate functional differentiation: as execution demands rise, monitoring responsibilities redistribute, supporting global performance, and preventing breakdowns in coordination. This network-analytic approach advances our understanding of team cognition in acute care. It showcases the utility of VR-based analytics for pinpointing phase-linked vulnerabilities and informing context-sensitive training.

For future work, we plan to extend our analyses beyond gaze data by integrating modalities such as team communication and physiological signals. We will also focus on event-centered analysis to capture how attention patterns reorganize around pivotal moments. Ultimately, these efforts aim to support richer modeling of individual and team situational awareness, guide training interventions, and enable real-time feedback for high-stakes collaborative environments.

\section{Declaration on Generative AI}
The authors used a HIPAA-compliant GPT tool and Grammarly for grammar and spelling checks. All content was subsequently reviewed and edited by the authors, who take full responsibility for its accuracy.

\bibliography{mTEAM_TNA}

\end{document}